\documentclass[twocolumn,preprintnumbers,amsmath,amssymb,prb,aps]{revtex4}
\usepackage{txfonts}
\usepackage{pifont}
\usepackage{amssymb}
\usepackage{dcolumn}
\usepackage{amsmath}
\usepackage[dvips]{epsfig}

\makeatletter
\def\btt#1{\texttt{\@backslashchar#1}}%
\DeclareRobustCommand\bblash{\btt{\@backslashchar}}%
\makeatother

\begin{document}
\title{Ground-state $\And$ finite-temperature properties of spin liquid phase in the $J_1-J_2$ honeycomb model}
\author{Xiang-Long Yu$^{1,2}$}
\author{Da-Yong Liu$^{1}$}
\author{Peng Li$^{3}$}
\author{Liang-Jian Zou$^{1}$}
\email[Corresponding author. E-mail: ]{zou@theory.issp.ac.cn}
\affiliation{1 Key Laboratory of Materials Physics,
              Institute of Solid State Physics, Chinese Academy of Sciences,
               P. O. Box 1129, Hefei 230031, China}
\affiliation{2 Graduate School of Chinese Academy of Sciences,
               Beijing 100000, China}
\affiliation{3 Center for Theoretical Physics, Department of Physics,
               Sichuan University, Chengdu 610064, China}
\date{Jan. 9, 2013}

\begin{abstract}
In this paper we analyze the groundstate and finite-temperature properties
of a frustrated Heisenberg $J_1-J_2$ model on a honeycomb lattice by employing
the Schwinger boson technique. The phase diagram and spin gap as functions of
${J}_{2}/{J}_{1}$ are presented, showing that the exotic spin liquid phase lies in
$0.21<{J}_{2}/{J}_{1} <0.43$. The temperature and magnetic-field dependences of
specific heat, magnetic susceptibility and Knight shift are also presented.
We find the spin liquid state is robust with respect to external magnetic field.
These results provide clear
information characterizing unusual properties of the exotic spin liquid phase
for further experiments.

\end{abstract}

\maketitle

\section{INTRODUCTION}

Frustrated quantum antiferromagnetic systems in low-dimensional lattices
have received great attention in recent years, especially on a honeycomb lattice.
Since they have a small coordination number $(z=3)$ in two dimension,
large quantum fluctuations are expected to lead to novel magnetic behaviors.
Some magnetic materials with honeycomb lattice were synthesized in recent
years and exhibit unusual properties. A recent high-field electron spin
resonance (ESR) spectroscopy measurement on novel
Bi$_{3}$Mn$_{4}$O$_{12}$(NO$_{3}$), which is a model substance of $S=3/2$
honeycomb lattice antiferromagnet (AFM), suggested an important role of the
geometric frustration in removing the long-range magnetic order, since no
long-range order was observed down to $1.9 K$ \cite{JPCS200.22042}.
In addition, a recent synthesized honeycomb compound In$_{3}$Cu$_{2}$VO$_{9}$
also exists weak intralayer and strong interlayer magnetic frustrations \cite{PRB85.085102,arXiv:1202.1861}, in which the spin-1/2 Cu sites forming a
honeycomb lattice do not exhibit a long-range magnetic order down to 2 K.
Furthermore, in the family of compounds BaM$_{2}$(XO$_{4}$)$_{2}$ with M=Co,
Ni and X=P, Regnault and Rossat-Mignod found that the magnetic ions $M$ with
spin $S=1/2$ for Co or $S=1$ for Ni form frustrated honeycomb layers. These
layers separate so large from each other that BaM$_{2}$(XO$_{4}$)$_{2}$
can be viewed as a two-dimensional AFM \cite{book1990phase,JPCS340.012012}.
The properties of these low-dimensional AFMs arise a strong conjecture that
whether they form a long-searched exotic spin liquid state.

Frustrated quantum magnets on honeycomb lattice may also exhibit remarkable
properties. As shown by Meng $et\ al.$\cite{NATURE464.847}, a quantum
spin liquid, which is a short-range resonating valence-bond (RVB) liquid due to
the frustration induced by the motion of the carriers, may be stable in an
intermediate-interacting Hubbard model on the honeycomb lattice, though their
results were questioned by Sorella $et\ al.$\cite{arXiv:1207.1783}.
Meanwhile, most of theoretical analysis focus on strongly interacting limit of
the Hubbard model, {\it i.e.} the Mott insulating region described by the
${J}_{1}-{J}_{2}$ AFM Heisenberg model on the honeycomb lattice.
Soon afterward Meng {\it et al.}'s work, Clark $et\ al.$ showed that in a
quantum $J_{1}-J_{2}$ model a spin liquid phase can be stable in the
range of $0.08 \leq J_{2}/J_{1} \leq 0.3$ \cite{PRL107.087204}.
Although the variational Monte Carlo approach based on the variational family
of entangled-plaquette states also supported that a spin liquid ground state is
stable in the range 0.2 $\leq J_{2}/J_{1} \leq$ 0.4 \cite{PRB85.060402}, many
other authors suggested a plaquette valence bond phase in the intermediate
$J_{2}/J_{1}$ ratio by the renormalization group method \cite{PRB84.14417},
the coupled-cluster method \cite{JPCM24.236002}
and the exact diagonalization \cite{PRB84.024406, JPCM23.226006, PRB84.012403}.
The divergency of zero-temperature magnetic phase diagrams on honeycomb lattice
obtained by different methods shows the necessarity of further study to clarify
these controversial results.

Meanwhile, very few work has come to the finite-temperature properties of
the frustrated honeycomb lattice though some have been studied in other structures \cite{JPSJ58.3733, JPSJ60.614, PRB60.1057, PRB43.7891, PRB38.316, PRB.40.5028, PRB69.064426, PRL61.617}. Such theoretical results could provide
the experimentists with some theoretical guidance to search for
exotic spin liquid phase in realistic materials.
For this purpose, we employ the Schwinger-boson mean-field theory (SBMFT), which has proved successful in incorporating quantum fluctuation, to investigate the finite-temperature properties on the frustrated honeycomb lattice. With respect to linear spin-wave theory, SBMFT can describe both ordered and disordered phases even with large fluctuation. Motivated by previous results, in this paper we focus solely on antiferromagnetic interactions ${J}_{ij}>0$ and present the thermodynamic properties by using SBMFT of the ${J}_{1}-{J}_{2}$ Heisenberg model for spin $S=1/2$ case.

We firstly find that with the increase of ${J}_{2}/{J}_{1}$, the system undergoes a transition from a N\'{e}el AFM phase to a quantum disordered phase at ${J}_{2}/{J}_{1}=0.21$; the disordered phase is classified as a spin liquid state, and sequently the disorder phase enters into a collinear striped AFM (SAFM) phase at the critical point ${J}_{2}/{J}_{1}=0.43$ at $T=0K$. We also clearly show that the spin liquid state is robust with respect to strong external magnetic field; and its low-temperature specific heat, magnetic susceptibility and Knight shift exhibit exponential behavior.
The rest of this paper is organized as follows: we first describe the model Hamiltonian and Schwinger boson method in Sec.II; for the theoretical and numerical results, we present the groundstate properties and zero-temperature phase diagram of the ${J}_{1}-{J}_{2}$ Heisenberg model in Sec. III and the finite-temperature properties of spin liquid phase in Sec. IV, respectively, and the final section is devoted to conclusion.

\section{Model Hamiltonian $\And$ Method}

\subsection{N\'{e}el AFM and Disordered Phase}

\begin{figure}[htbp]
\centering
\setlength{\abovecaptionskip}{2pt}
\setlength{\belowcaptionskip}{4pt}
\includegraphics[angle=0, width=0.8 \columnwidth]{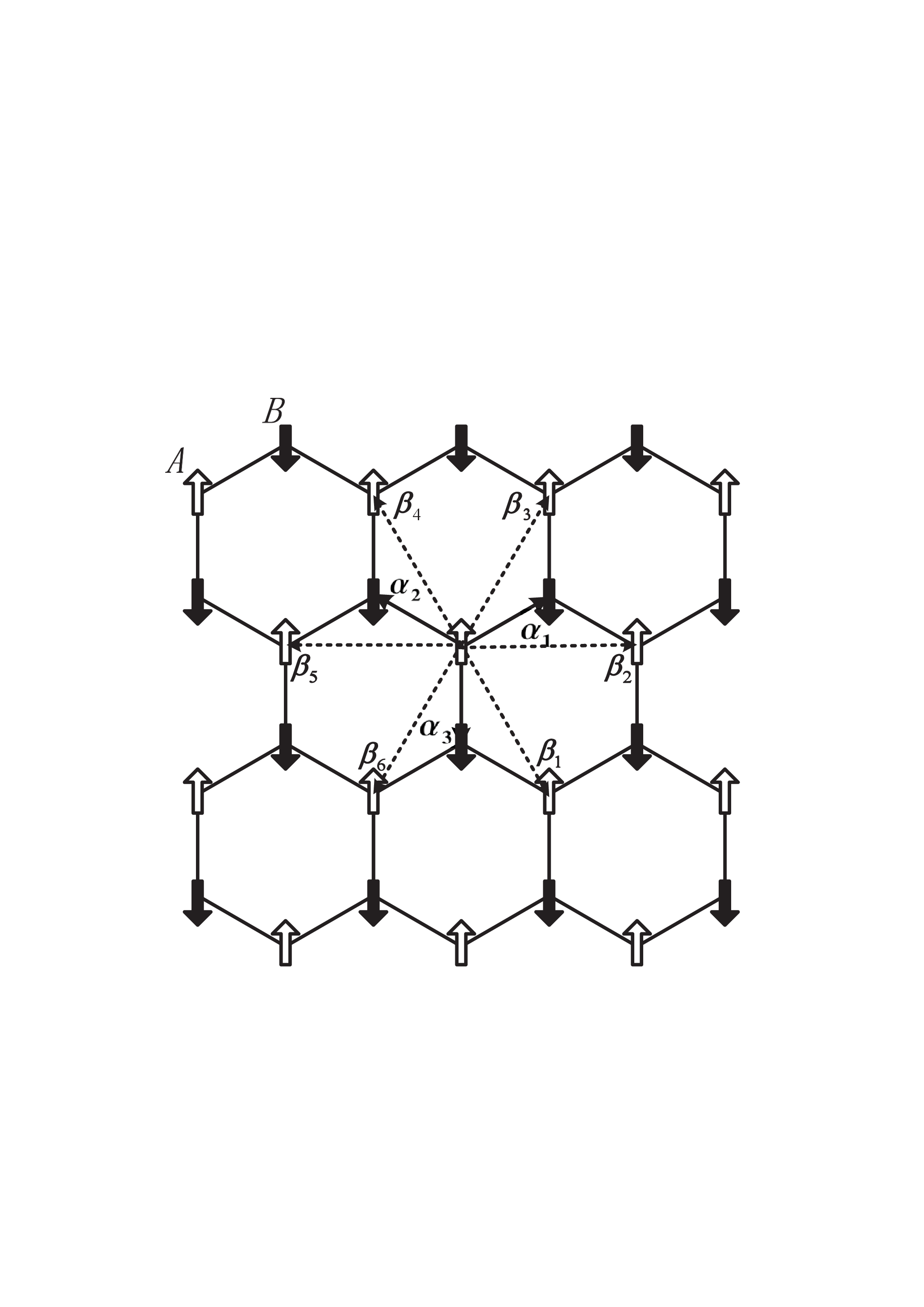}
\caption{N\'{e}el AFM honeycomb lattice with sublattices A and B. The vectors $\alpha $ and $\beta $ point to the nearest neighbors and next-nearest neighbors of a central site, respectively.}
\label{fig:NEEL honeycomb lattice}
\end{figure}

The honeycomb lattice is composed of two interlacing triangular lattices. Every site has three nearest neighbors on the other lattice, and six next-nearest neighbors on the same lattice. The ${J}_{1}-{J}_{2}$ model Hamiltonian reads
\begin{eqnarray}
H={{J}_{1}}\sum\limits_{r,\alpha }{{{S}_{r}}\cdot {{S}_{r+\alpha }}}+{{J}_{2}}\sum\limits_{R,\beta }{{{S}_{R}}\cdot {{S}_{R+\beta }}}.
\end{eqnarray}
Here ${J}_{1}$ and ${J}_{2}$ are the nearest-neighbor and next-nearest-neighbor AFM couplings, respectively. This system is frustrated when both ${J}_{1}$ and ${J}_{2}$ are positive. The ${\alpha }$ and ${\beta }$ sums run over all three nearest-neighbor vectors ${\alpha }$ of site ${r}$ and the six next-nearest-neighbor vectors ${\beta }$ of site ${R}$. The vectors describing the neighbor positions are shown in Fig. \ref{fig:NEEL honeycomb lattice}. Let the lattice constant equal to unity.\\

Since the honeycomb lattice is a complex Bravais lattice, we separately treat the sublattice spins with different Schwinger bosons. The spin operators are represented as follows:
\begin{align}
  & S_{i}^{z}=\frac{1}{2}\left( a_{i\uparrow }^{+}{{a}_{i\uparrow }}-a_{i\downarrow }^{+}{{a}_{i\downarrow }} \right) \nonumber, \\
 & \ S_{i}^{+}=a_{i\uparrow }^{+}{{a}_{i\downarrow }},\ S_{i}^{-}=a_{i\downarrow }^{+}{{a}_{i\uparrow }}, \\ \nonumber
\end{align}
for sublattice A. The spins on the sublattice B are rotated by $\pi$ from those on the sublattice B.
\begin{equation}
{{a}_{i\uparrow }}\to -{{b}_{j\downarrow }},\ {{a}_{i\downarrow }}\to {{b}_{j\uparrow }}.
\label{eq:rotation}
\end{equation}
Within the Schwinger boson representation, the sublattice interactions are rewritten as
\begin{align}
  &
  {{H}_{1}}=-\frac{1}{2}{{J}_{1}}\sum\limits_{<i,j>}{:A_{ij}^{+}{{A}_{ij}}:}+
  \frac{3}{2}N{{J}_{1}}{{S}^{2}}, \\
 & {{H}_{2}}=\frac{1}{2}{{J}_{2}}\sum\limits_{<i,l>}{:B_{il}^{+}{{B}_{il}}:}-\frac{3}{2}N{{J}_{2}}{{S}^{2}}, \\
 & {{H}_{3}}=\frac{1}{2}{{J}_{2}}\sum\limits_{<j,k>}{:C_{jk}^{+}{{C}_{jk}}:}-\frac{3}{2}N{{J}_{2}}{{S}^{2}},
\end{align}
where ${{H}_{1}}$ is the Hamiltonian of the interaction between the two different sublattices, ${{H}_{2}}$ is that of the sublattice A, and ${{H}_{3}}$ is that of the sublattice B; the operators ${{A}_{ij}}={{a}_{i\uparrow }}{{b}_{j\uparrow }}+{{a}_{i\downarrow }}{{b}_{j\downarrow }},\ {{B}_{il}}=a_{i\uparrow }^{+}{{a}_{l\uparrow }}+a_{i\downarrow }^{+}{{a}_{l\downarrow }}$ and ${{C}_{jk}}=b_{j\uparrow }^{+}{{b}_{k\uparrow }}+b_{j\downarrow }^{+}{{b}_{k\downarrow }}$ correspond to AFM correlation of nearest neighbors and ferromagnetic (FM) correlation of next-nearest neighbors in the sublattice A and B, respectively. $N$ is the number of all lattice sites. The total Hamiltonian then reads:
\begin{align}
{H}=&{{H}_{1}}+{{H}_{2}}+{{H}_{3}}+\sum\limits_{i}{{{\lambda }_{i}}\left( a_{i\uparrow }^{+}{{a}_{i\uparrow }}+a_{i\downarrow }^{+}{{a}_{i\downarrow }}-2S \right)} \nonumber \\ &+\sum\limits_{j}{{{\lambda }_{j}}\left( b_{j\uparrow }^{+}{{b}_{j\uparrow }}+b_{j\downarrow }^{+}{{b}_{j\downarrow }}-2S \right)},
\end{align}
where the local constraints for the spin Hilbert space are enforced with the Lagrange multipliers ${{\lambda }_{i}}$ and ${{\lambda }_{j}}$, which will be replaced by a local parameter ${\lambda }$ in our mean-field treatment.
Assuming the averages of order parameters of the boson fields $P=<{{A}_{ij}}>$ and $Q=<{{B}_{il}}>=<{{C}_{jk}}>$, we decouple the Hamiltonian into quadratic, and the effective $H$ can be diagonalized.

We use a complex Bogoliubov transformation in the momentum space
\begin{align}
  & {{a}_{k\sigma }}=u_{k}^{*}{{\alpha }_{k\sigma }}+{{v}_{k}}\beta _{k\sigma }^{+}, \
  {{b}_{k\sigma }}={{v}_{k}}\alpha _{k\sigma }^{+}+u_{k}^{*}{{\beta }_{k\sigma }}
\end{align}
to diagonalize the Halmiltonian by choosing $u_{k}$ and $v_{k}$ properly. The diagonalized Hamiltonian
\begin{equation}
H=\sum\limits_{k\sigma }{[{{\omega }_{k}}(\alpha _{k\sigma }^{+}{{\alpha }_{k\sigma }}+\beta _{k\sigma }^{+}{{\beta }_{k\sigma }}+1)]}+{{E}_{C}}
\end{equation}
has the energy dispersion relation of the Schwinger boson excitation ${\omega }_{k}$
\begin{equation}
\label{eq:dispersion relationship}
{{\omega }_{k}}=\sqrt{X_{k}^{2}-Y_{k}^{2}}, \
\end{equation}
and the constant energy
\begin{equation}
\frac{{{E}_{C}}}{N}=\frac{3}{4}{{J}_{1}}{{P}^{2}}-\frac{3}{2}{{J}_{2}}Q{}^{2}+
\frac{3}{2}{{J}_{1}}{{S}^{2}}-3{{J}_{2}}{{S}^{2}}-\lambda -2S\lambda
\end{equation}
 and
\begin{align}
{{X}_{k}}=\lambda +3{{J}_{2}}Q{{r}_{2k}}, \ {{Y}_{k}}=\frac{3}{2}{{J}_{1}}\left|P{{r}_{1k}} \right|
\end{align}
where the geometrical structure factors are given by
\begin{align}
  & {{\gamma }_{1k}}=\frac{1}{3}\sum\limits_{\alpha }{{{e}^{-ik\alpha }}}, \ {{\gamma }_{2k}}=\frac{1}{6}\sum\limits_{\beta }{\cos (k\cdot \beta )}.
\end{align}
Thus minimizing the free energy $F=-{{k}_{B}}T\ln Tr\left( {{e}^{-\beta H}} \right)$ with respect to $P, Q$ and $\lambda$, we arrive a set of self-consistent equations
\begin{align}
  & \frac{2}{N}\sum\limits_{k}{\left[ \frac{{{X}_{k}}}{{{\omega }_{k}}}\left( 2{{n}_{k}}+1 \right) \right]}-\left( 2S+1 \right)=0,
 \label{eq:s-c eq1} \\
 & \frac{2}{N}\sum\limits_{k}{\left[ \frac{\left| {{r}_{1k}} \right|{{Y}_{k}}}{{{\omega }_{k}}}\left( 2{{n}_{k}}+1 \right) \right]}-P=0,
 \label{eq:s-c eq2} \\
 & \frac{2}{N}\sum\limits_{k}{\left[ \frac{{{r}_{2k}}{{X}_{k}}}{{{\omega }_{k}}}\left( 2{{n}_{k}}+1 \right) \right]}-Q=0, \label{eq:s-c eq3}
\end{align}
with ${{n}_{k}}=1/[\exp (\beta {{\omega }_{k}})-1]$ being the Bose-Einstein distribution function and
 $\beta ={1}/{{{k}_{B}}T}$.
Solving the mean-field equations yields the average values of $P$, $Q$, and $\lambda$, we can explore properties of this system at zero and finite temperatures.

In a two-dimensional honeycomb system, an AFM long-range order (LRO) can be stable when $J_{2}$ is small enough at low temperature. As the AFM LRO corresponds to a condensation of the Schwinger bosons, we could obtain the self-consistent equations at $T = 0$ K by the replacement
\begin{align}
\frac{{{n}_{k}}}{{{\omega }_{k}}}\to {{S}^{*}}\delta (k), \
\end{align}
where ${{S}^{*}}$ measures the fraction of the Bose-Einstein condensation (BEC) of the Schwinger bosons \cite{PRB43.7891}.
When this system is a LRO corresponding to BEC at $T = 0$ K, the ground-state energy per site reads:
\begin{equation}
{{\varepsilon }_{0}}=\frac{2}{N}\sum\limits_{k}{{{\omega }_{k}}}+\frac{{{E}_{c}}}{N}.
\label{eq:ground sate energy}
\end{equation}
On the other hand, the system does not undergo the BEC due to frustration effect when $J_{2}/J_{1}$ becomes large. We could separate the BEC phase with AFM LRO from the non-BEC phase with spin disorder by judging the vanishing ${{S}^{*}}$ and gap for $S=1/2$\cite{PRB43.7891}. The ground-state energy per site of the non-BEC phase, corresponding to the spin disorder phase, is the same to Eq. (\ref{eq:ground sate energy}).

\subsection{Collinear Striped Antiferromagnetic Phase}

\begin{figure}[htbp]
\centering
\setlength{\abovecaptionskip}{2pt}
\setlength{\belowcaptionskip}{4pt}
\includegraphics[angle=0, width=0.8 \columnwidth]{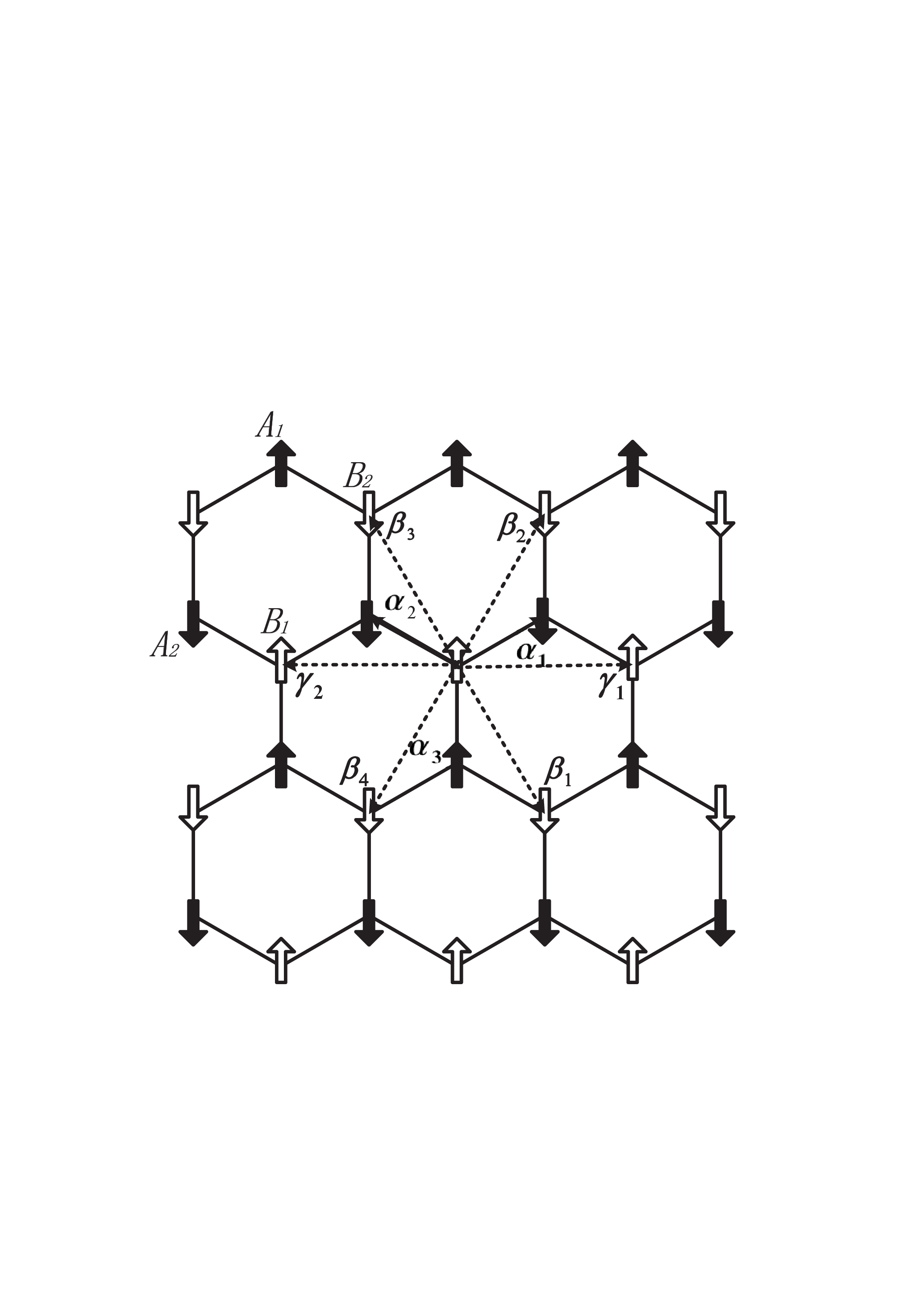}
\caption{SAFM sublattices ${A}_{1}, {A}_{2}, {B}_{1}$ and ${B}_{2}$ in honeycomb lattice. The vectors $\alpha $, $\beta $ and $\gamma $ point to the nearest and next-nearest neighbors.}
\label{fig:collinear honeycomb lattice}
\end{figure}
When ${J}_{2}/{J}_{1}$ is large enough, from the classical energy analysis, one finds that the spin disorder phase becomes unstable, and the collinear SAFM phase becomes more stable.  The magnetic structure and the vectors describing the neighbor positions are shown in Fig. \ref{fig:collinear honeycomb lattice}.

The SAFM honeycomb lattice composes of four interacting sublattices (${A}_{1}$, ${A}_{2}$, ${B}_{1}$ and ${B}_{2}$), corresponding to four Schwinger bosons. After a rotation of $\pi$ in spin space in ${A}_{2}$ and ${B}_{2}$ sites with respect to that in ${A}_{1}$ and ${B}_{1}$ sites, similar to Eq. (\ref{eq:rotation}), the spin operators in ${A}_{1}$ and ${A}_{2}$ sublattices can be represented as follows:
\begin{align}
  & S_{i1}^{z}=\frac{1}{2}\left( a_{i1\uparrow }^{+}{{a}_{i1\uparrow }}-a_{i1\downarrow }^{+}{{a}_{i1\downarrow }} \right),\nonumber \\
  & S_{i1}^{+}=a_{i1\uparrow }^{+}{{a}_{i1\downarrow }},\  S_{i1}^{-}=a_{i1\downarrow }^{+}{{a}_{i1\uparrow }};\  \\
 & S_{i2}^{z}=-\frac{1}{2}\left( a_{i2\uparrow }^{+}{{a}_{i2\uparrow }}-a_{i2\downarrow }^{+}{{a}_{i2\downarrow }} \right),\nonumber \\
 & S_{i2}^{+}=-a_{i2\downarrow }^{+}{{a}_{i2\uparrow }},\  S_{i2}^{-}=-a_{i2\uparrow }^{+}{{a}_{i2\downarrow }}.\
\end{align}
Replacing operator $a$ with $b$ gives rise to the Schwinger representation for ${B}_{1}$ and ${B}_{2}$ sublattices.
Rewriting $H$ and making the mean-field approximation similar to that in the N\'{e}el AFM phase,
we get an effective Hamiltonian:\\
\begin{align}
   H=\sum\limits_{k\sigma }&{ \left( \begin{matrix}
   a_{k,1\sigma }^{+} & b_{k,1\sigma }^{+} & {{a}_{k,2\sigma }} & {{b}_{k,2\sigma }}  \\
\end{matrix} \right)\left( \begin{matrix}
   A & {{C}^{*}} & {{D}^{*}} & B  \\
   C & A & B & D  \\
   D & B & A & C  \\
   B & {{D}^{*}} & {{C}^{*}} & A  \\
\end{matrix} \right)\left( \begin{matrix}
   {{a}_{k,1\sigma }}  \\
   {{b}_{k,1\sigma }}  \\
   a_{k,2\sigma }^{+}  \\
   b_{k,2\sigma }^{+}  \\
\end{matrix} \right) } \nonumber \\
&+{{E}_{C}}
\end{align}
with the constant energy
\begin{align}
 \frac{{{E}_{C}}}{N}=\frac{1}{2}{{J}_{1}}P_{1}^{2}+{{J}_{2}}P_{2}^{2}-\frac{1}{4}{{J}_{1}}Q_{1}^{2}
 -\frac{1}{2}{{J}_{2}}Q_{2}^{2} \nonumber \\ +\frac{1}{2}{{J}_{1}}{{S}^{2}}+{{J}_{2}}{{S}^{2}}-\lambda-2S\lambda
\end{align}
where we define
\begin{align}
 &A={{J}_{2}}{{Q}_{2}}{{r}_{1k}}+\lambda ,\ \nonumber
 B=-2{{J}_{2}}{{P}_{2}}{{r}_{2k}},\nonumber \\
 &C=\frac{1}{2}{{J}_{1}}{{Q}_{1}}{{r}_{3k}},\
 D=-{{J}_{1}}{{P}_{1}}{{r}_{4k}};
\end{align}
 and the geometrical structure factors are
\begin{align}
  &{{r}_{1k}}=\frac{1}{2}\sum\limits_{_\gamma}{\cos \left( k\cdot \gamma  \right)}, \ \nonumber
  {{r}_{2k}}=\frac{1}{4}\sum\limits_{\beta }{{{e}^{ik\beta }}},\nonumber \\
 &{{r}_{3k}}={{e}^{ik{{\alpha }_{3}}}},\
  {{r}_{4k}}=\frac{1}{2}\sum\limits_{{{\alpha }_{1,2}}}{{{e}^{ik{{\alpha }_{1,2}}}}}.
\end{align}
${{P}_{1}},\ {{P}_{2}},\ {{Q}_{1}}$ and ${{Q}_{2}}$ are the order parameters, corresponding to the nearest-neighbor AFM coupling, next-nearest-neighbor AFM coupling, nearest-neighbor FM coupling, and next-nearest-neighbor FM coupling, respectively, and $\lambda$ is the local Lagrange multiplier.After a complex $4\times 4$ Bogoliubov transformation to diagonalize the Halmiltonian\cite{PR2.A450}, we get the following self-consistent equations by minimizing the free energy $F=-{{k}_{B}}T\ln Tr\left( {{e}^{-\beta H}} \right)$:
\begin{align}
  & \frac{4}{N}\sum\limits_{k}{\left[ \left( {{n}_{k1}}+\frac{1}{2} \right)\frac{\partial {{\omega }_{k1}}}{\partial \lambda }+\left( {{n}_{k2}}+\frac{1}{2} \right)\frac{\partial {{\omega }_{k2}}}{\partial \lambda } \right]}-\left( 2S+1 \right)=0,
\end{align}
\begin{align}
 & \frac{4}{N}\sum\limits_{k}{\left[ \left( {{n}_{k1}}+\frac{1}{2} \right)\frac{\partial {{\omega }_{k1}}}{\partial {{P}_{1}}}+\left( {{n}_{k2}}+\frac{1}{2} \right)\frac{\partial {{\omega }_{k2}}}{\partial {{P}_{1}}} \right]}+{{J}_{1}}{{P}_{1}}=0,
\end{align}
\begin{align}
 & \frac{4}{N}\sum\limits_{k}{\left[ \left( {{n}_{k1}}+\frac{1}{2} \right)\frac{\partial {{\omega }_{k1}}}{\partial {{P}_{2}}}+\left( {{n}_{k2}}+\frac{1}{2} \right)\frac{\partial {{\omega }_{k2}}}{\partial {{P}_{2}}} \right]}+2{{J}_{2}}{{P}_{2}}=0,
\end{align}
\begin{align}
 & \frac{4}{N}\sum\limits_{k}{\left[ \left( {{n}_{k1}}+\frac{1}{2} \right)\frac{\partial {{\omega }_{k1}}}{\partial {{Q}_{1}}}+\left( {{n}_{k2}}+\frac{1}{2} \right)\frac{\partial {{\omega }_{k2}}}{\partial {{Q}_{1}}} \right]}-\frac{1}{2}{{J}_{1}}{{Q}_{1}}=0,
\end{align}
\begin{align}
 & \frac{4}{N}\sum\limits_{k}{\left[ \left( {{n}_{k1}}+\frac{1}{2} \right)\frac{\partial {{\omega }_{k1}}}{\partial {{Q}_{2}}}+\left( {{n}_{k2}}+\frac{1}{2} \right)\frac{\partial {{\omega }_{k2}}}{\partial {{Q}_{2}}} \right]}-{{J}_{2}}{{Q}_{2}}=0,
\end{align}
with the Bose-Einstein distribution function ${{n}_{k1,2}}=1/[\exp (\beta {{\omega }_{k1,2}})-1]$ and the dispersion relations
\begin{widetext}
\begin{equation}
{{\omega }_{k1,2}}=\sqrt{{{A}^{2}}-{{B}^{2}}+{{\left| C \right|}^{2}}-{{\left| D \right|}^{2}}\pm \sqrt{4{{A}^{2}}{{\left| C \right|}^{2}}+4{{B}^{2}}{{\left| D \right|}^{2}}-4\left( AB{{C}^{*}}D+ABC{{D}^{*}} \right)+{{\left( C{{D}^{*}}-{{C}^{*}}D \right)}^{2}}}}
\end{equation}
\end{widetext}
Since this LRO state is characterized by a BEC of bosons at ${k}_{0}=(0, 0)$ in the Schwinger boson approach, a straightforward calculation yields the ground-state energy per site in the collinear SAFM phase
\begin{equation}
{{\varepsilon }_{0}}=\frac{2}{N}\sum\limits_{k}{\left( {{\omega }_{k1}}+{{\omega }_{k2}} \right)}+\frac{{{E}_{C}}}{N}.
\end{equation}
Thus, we could compare the ground-state energies among N\'{e}el AFM, spin liquid (disorder) and SAFM phase, and pick out the stablest ground state.

\subsection{Spin-Spin Correlation Functions}

We could also understand the spin structure of the different phases by exploring the spin-spin correlation functions (SSCF). In the presence of the long-range N\'{e}el AFM order, the SSCF consists of the transverse and longitudinal correlations\cite{PRB60.1057}.
The longitudinal correlation function between different sublattice sites is expressed as
\begin{align}
  & \left\langle S_{i}^{z}\cdot S_{j}^{z} \right\rangle =\left\langle -\frac{1}{4}\left( a_{i\uparrow }^{+}{{a}_{i\uparrow }}-a_{i\downarrow }^{+}{{a}_{i\downarrow }} \right)\left( b_{j\uparrow }^{+}{{b}_{j\uparrow }}-b_{j\downarrow }^{+}{{b}_{j\downarrow }} \right) \right\rangle  \nonumber \\
 & \ \ \ \ \ \ \ \ \ \ \ \ \ \ \ =-2{{\left( 1-{{M}_{1}} \right)}^{2}}-\frac{1}{2}M_{2}^{2}-\left( 1-{{M}_{1}} \right){{M}_{2}},
 \label{eq:longitudinal cf}
\end{align}
 where
 \begin{align}
& {{M}_{1}}=\frac{2}{N}\sum\limits_{k}{\frac{{{X}_{k}}}{2{{\omega }_{k}}}}, \\
& {{M}_{2}}= \frac{2}{N}\sum\limits_{k}\left\{ \left[ \cos \left( \overrightarrow{k}\cdot \overrightarrow{{{R}_{ij}}} \right)\left( r_{1k}^{*}+{{r}_{1k}} \right) \right.\right. \nonumber \\
 &\ \ \ \ \ \ \ \ \ \ \ \ \ \ \left.\left. +i\cdot \sin \left( \overrightarrow{k}\cdot \overrightarrow{{{R}_{ij}}} \right)\left( r_{1k}^{*}-{{r}_{1k}} \right) \right]\left( \frac{3{{J}_{1}}P}{8{{\omega }_{k}}} \right) \right\}.
\end{align}
The average to various paired terms with four operators in Eq. (\ref{eq:longitudinal cf}) is in accordance with Wick theorem.\\
The transverse correlation functions between sublattice sites read
\begin{align}
  & \left\langle S_{i}^{x}\cdot S_{j}^{x} \right\rangle =\left\langle S_{i}^{y}\cdot S_{j}^{y} \right\rangle =-2{{M}_{2}}+2{{M}_{1}}{{M}_{2}}-M_{2}^{2}
\end{align}
Then in the presence of LRO the SSCFs between the sites belonging to the same lattice are calculated,
\begin{align}
  & \left\langle S_{i}^{z}\cdot S_{l}^{z} \right\rangle =2{{\left( 1-{{N}_{1}} \right)}^{2}}+\left( 1-{{N}_{1}} \right){{N}_{2}}+\frac{1}{2}N_{2}^{2}, \\
 & \left\langle S_{i}^{+}\cdot S_{l}^{-} \right\rangle =2{{N}_{2}}-2{{N}_{1}}+N_{2}^{2},
\end{align}
 where \\
 \begin{align}
 & {{N}_{1}}=\frac{2}{N}\sum\limits_{k}{\frac{{{X}_{k}}}{2{{\omega }_{k}}}}, \\
 & {{N}_{2}}=\frac{2}{N}\sum\limits_{k}{\cos \left( \overrightarrow{k}\cdot \overrightarrow{{{R}_{il}}} \right)\frac{{{X}_{k}}}{2{{\omega }_{k}}}}.
\end{align}
In a similar way we can get the SSCF in the spin liquid phase, where the three components of the correlation functions are all equal.
The SSCF between different sublattice sites thus reads
\begin{align}
\left\langle \overrightarrow{{{S}_{i}}}\cdot \overrightarrow{{{S}_{j}}} \right\rangle =-\frac{3}{2} & \left\{  \frac{2}{N}\cdot \sum\limits_{k}\left[ \cos \left( \overrightarrow{k}\cdot \overrightarrow{{{R}_{ij}}} \right)\left( r_{1k}^{*}+{{r}_{1k}} \right) \right.\right. \nonumber \\
& \left.\left. +i\cdot \sin \left( \overrightarrow{k}\cdot \overrightarrow{{{R}_{ij}}} \right)\left( r_{1k}^{*}-{{r}_{1k}} \right) \right]\left( \frac{3{{J}_{1}}P}{8{{\omega }_{k}}} \right) \right\}^{2}
\end{align}
The SSCF between different sites belonging to the same sublattice can be calculated as
\begin{align}
\left\langle \overrightarrow{{{S}_{i}}}\cdot \overrightarrow{{{S}_{j}}} \right\rangle =\frac{3}{2} \times & \frac{2}{N}\sum\limits_{k}{\left\{ \cos \left( \overrightarrow{k}\cdot \overrightarrow{{{R}_{il}}} \right)\left[ \frac{3{{J}_{1}}P\left| {{r}_{1k}} \right|}{4{{\omega }_{k}}}-\frac{1}{2} \right] \right\}} \nonumber \\
& \times \frac{2}{N}\sum\limits_{{{k}^{'}}}{\left\{ \cos \left( \overrightarrow{{{k}^{'}}}\cdot \overrightarrow{{{R}_{il}}} \right)\left[ \frac{3{{J}_{1}}P\left| {{r}_{1{{k}^{'}}}} \right|}{4{{\omega }_{{{k}^{'}}}}}+\frac{1}{2} \right] \right\}}.
\end{align}
The SSCF of the collinear SAFM phase can also be obtained through calculating longitudinal and transverse correlation functions by using the method similar to that of N\'{e}el AFM.\\

\subsection{Magnetic Field Effect $\And$ Knight Shift}

When a magnetic field B is applied parallel to the z-axis, similar to the previous process, we can get the diagonalized Hamiltonian,
\begin{align}
   {{H}_{MF}}= & \sum\limits_{k} \left[ {{\omega }_{k1}}\left( \alpha _{k\uparrow }^{+}{{\alpha }_{k\uparrow }}+\beta _{k\downarrow }^{+}{{\beta }_{k\downarrow }} \right)  +{{\omega }_{k2}}\left( \alpha _{k\downarrow }^{+}{{\alpha }_{k\downarrow }}+\beta _{k\uparrow }^{+}{{\beta }_{k\uparrow }} \right)\right.\nonumber\\
&\ \ \ \ \ \ \ \ \ \ \ \ \ \ \ \ \ \ \ \ \ \ \ \ \ \ \ \ \ \ \ \ \ \ \ \ \ \ \ \ \ \ \left.  +2\sqrt{X_{k}^{2}-Y_{k}^{2}}\right]+{{E}_{C}}
\end{align}
where
\begin{align}
\label{eq:omega_B}
& {{\omega }_{k1,2}}={{\omega }_{k0}}\pm \frac{1}{2}\mu B, \\
& {{\omega }_{k0}}=\sqrt{X_{k}^{2}-Y_{k}^{2}}, \\
& {{X}_{k}}=\lambda +3{{J}_{2}}Q{{r}_{2k}}, \ {{Y}_{k}}=\frac{3}{2}{{J}_{1}}\left|P{{r}_{1k}} \right|,
\end{align}
$\mu =g{{\mu }_{B}}$, $g$ is the Lande factor, ${{\mu }_{B}}$ is the Bohr magneton
and the constant enegy\\
\begin{equation}
\frac{{{E}_{C}}}{N}=\frac{3}{4}{{J}_{1}}{{P}^{2}}-\frac{3}{2}{{J}_{2}}{{Q}^{2}}+
\frac{3}{2}{{J}_{1}}{{S}^{2}}-3{{J}_{2}}{{S}^{2}}-\lambda -2S\lambda. \\
\end{equation}
We get the self-consistent equations by minimizing the free energy:
\begin{align}
  & \frac{2}{N}\sum\limits_{k}{\left[ \frac{{{X}_{k}}}{{{\omega }_{k0}}}\left( {{n}_{k1}}+{{n}_{k2}}+1 \right) \right]}-\left( 2S+1 \right)=0, \\
 & \frac{2}{N}\sum\limits_{k}{\left[ \frac{\left| {{r}_{1k}} \right|{{Y}_{k}}}{{{\omega }_{k0}}}\left( {{n}_{k1}}+{{n}_{k2}}+1 \right) \right]}-P=0, \\
 & \frac{2}{N}\sum\limits_{k}{\left[ \frac{{{r}_{2k}}{{X}_{k}}}{{{\omega }_{k0}}}\left( {{n}_{k1}}+{{n}_{k2}}+1 \right) \right]}-Q=0,
\end{align}
which are reasonable generalization of Eq. (\ref{eq:s-c eq1}-\ref{eq:s-c eq3}). Here n$_{k1,2}$ represents the bosonic occupation with the spectrum $\omega_{k1,2}$.

Nuclear magnetic resonance(NMR) is a classical method for probing the magnetic correlation effects, and Knight shift can directly measure the spin fluctuations around a nucleus in NMR experiment\cite{PRL3.262,book-principles}.
When the magnetic field $B$ parallel to z-axis is not too large, the system remains in the spin liquid phase, the nuclear spin is polarized along this direction, and the energy level is split. Simultaneously, the electron spin is also polarized, and its polarization can impact on the nucleus through the hyperfine interaction. Accordingly, NMR frequency is changed.
The effective magnetic field generated by polarized electrons around nucleus is given as follows\cite{PhysC157.561,PRB40.11382},
\begin{equation}
\delta {{B}}=\frac{F\left( q \right)\left\langle {{S}_{z}}\left( q \right) \right\rangle {{\delta }_{q,0}}}{{{\gamma }_{N}}\hbar }
\end{equation}
where $F\left( q \right)$ is the hyperfine interaction structure factor, and ${{\gamma }_{N}}$ is the nuclear gyromagnetic ratio.
Thus, we can get the Knight-shift\cite{xiangtao},
\begin{equation}
K=\frac{\delta \omega }{\omega }=\frac{{{\gamma }_{N}}\cdot \delta {{B}}}{{{\gamma }_{N}}\cdot {{B}}}=\frac{F\left( 0 \right)}{{{\gamma }_{N}}\hbar {\mu }}\cdot \frac{M\left( T \right)}{{{B}}}, \\
\end{equation}
from which the local magnetic field could be characterized.

\section{Ground-State Properties and Phase Diagram}

\begin{figure}[htbp]
\centering
\setlength{\abovecaptionskip}{2pt}
\setlength{\belowcaptionskip}{4pt}
\includegraphics[angle=0, width=1.0 \columnwidth]{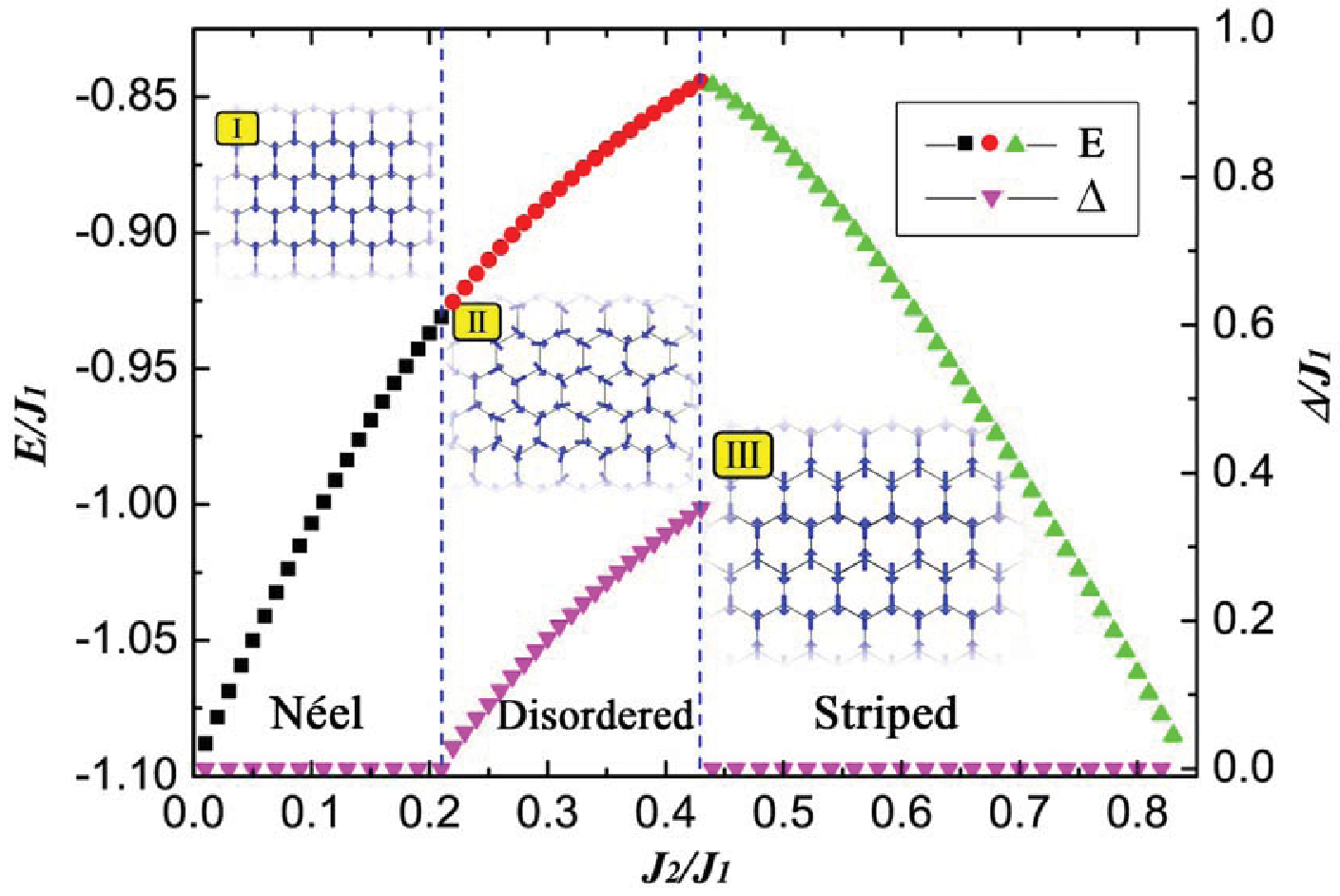}
\caption{(Color online) Ground-state energy per site and energy gap in the bosonic dispersion as a function of ${J}_{2}/{J}_{1}$ for $S=1/2$. For ${J}_{2}/{J}_{1}<0.21$ and ${J}_{2}/{J}_{1}>0.43$, the system remains gapless, corresponding to N\'{e}el AFM and collinear SAFM phase, respectively. When $0.21<{J}_{2}/{J}_{1}<0.43$ the system remains gapped.}
\label{fig:egap}
\end{figure}

We first present the zero-temperature phase diagram of ${J}_{2}/{J}_{1}$ by comparing the groundstate energy, as shown in Fig. \ref{fig:egap}. There are two critical points. The first one at $(J_{2}/J_{1})_{c1}=0.21$, which is in agreement with Mattsson's result\cite{PRB49.3997}, corresponds to a continuous phase transition between the N\'{e}el AFM and the spin disordered phases. The second one at $(J_{2}/J_{1})_{c2}=0.43$ corresponds to a first-order phase transition between the spin disordered phase and the collinear SAFM one. In comparison with the results by
Mezzacapo $et\ al.$, the range of our spin disordered phase ($0.21<{J}_{2}/{J}_{1}<0.43$) is almost in agreement with that obtained from variational Monte Carlo calculations (about 0.2 $\leq J_{2}/J_{1} \leq$ 0.4), based on the variational family of entangled-plaquette states \cite{PRB85.060402}.

\begin{figure}[htbp]
\centering
\setlength{\abovecaptionskip}{2pt}
\setlength{\belowcaptionskip}{4pt}
\includegraphics[angle=0, width=0.495 \columnwidth]{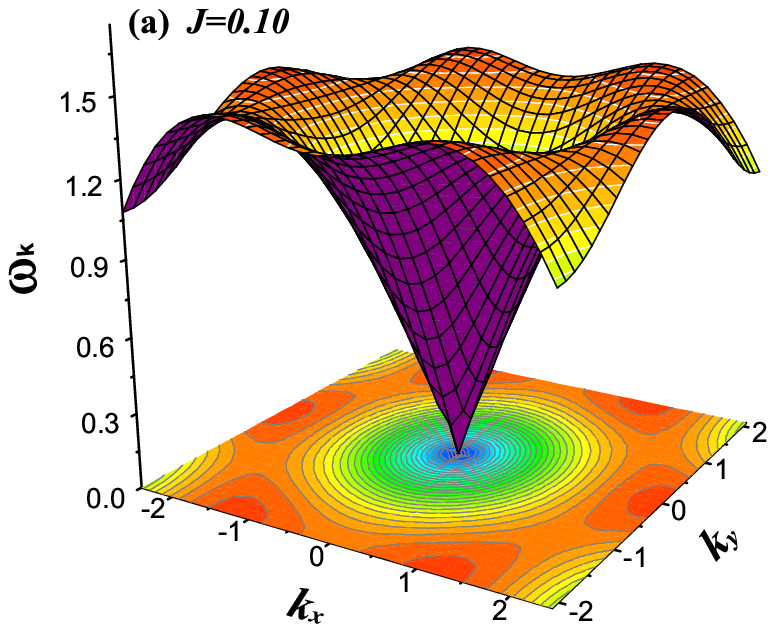}
\includegraphics[angle=0, width=0.495 \columnwidth]{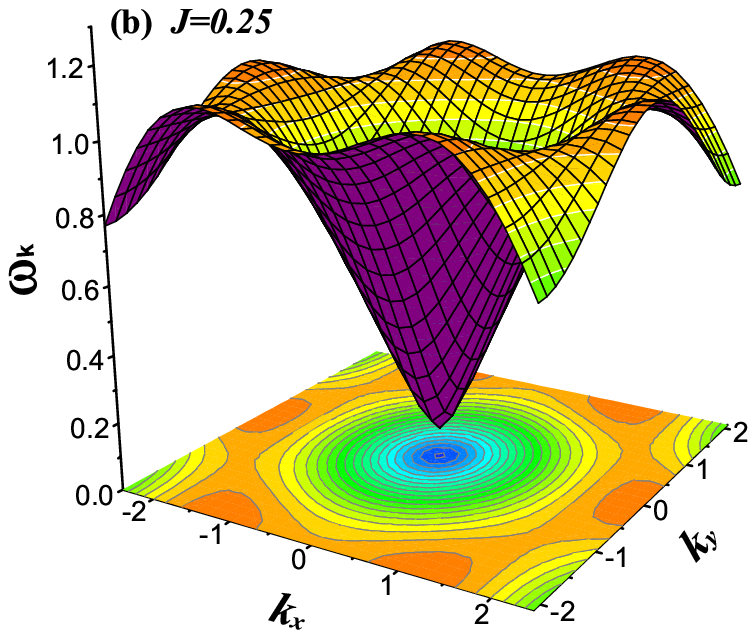}
\includegraphics[angle=0, width=0.495 \columnwidth]{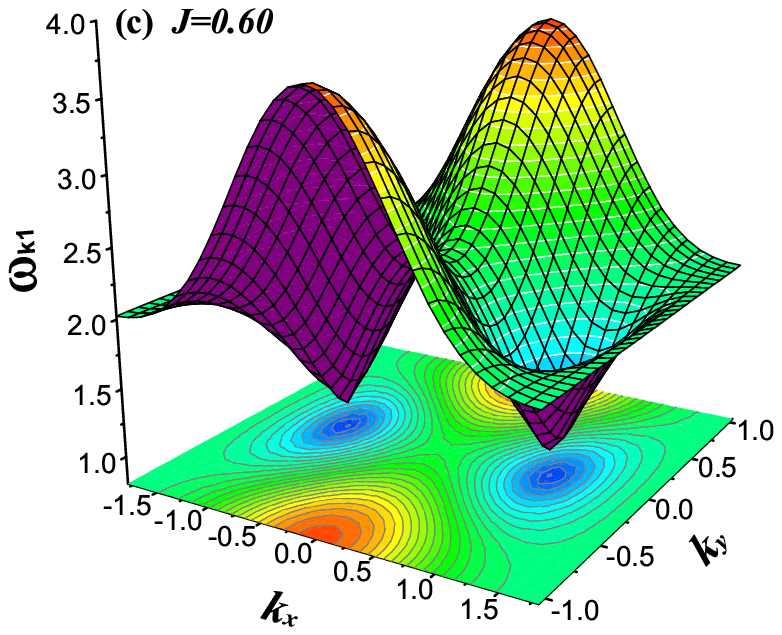}
\includegraphics[angle=0, width=0.495 \columnwidth]{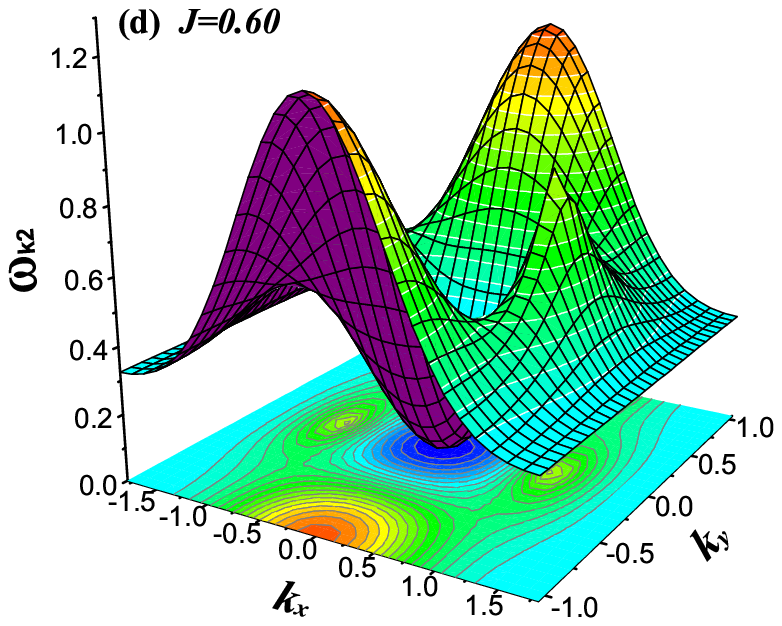}
\caption{(Color online) The dispersion relation for ${J}=0.10$ (a), $0.25$ (b), $0.60$ (c) and (d) at zero temperature. (c) and (d) correspond to two spin waves in the collinear SAFM phase. In the following figures, ${J}={J}_{2}/{J}_{1}$.}
\label{fig:wk}
\end{figure}

In accordance with the phase diagram, the energy gap of the Schwinger bosons as a function of ${J}_{2}/{J}_{1}$ is finite only in the spin disordered region $0.21<{J}_{2}/{J}_{1}<0.43$, and the gap lifts with the increase of ${J}_{2}/{J}_{1}$, shown in Fig. \ref{fig:egap}. The energy gaps vanish when ${J}_{2}/{J}_{1} < 0.21$ or $>0.43$, where the BEC and magnetic LRO appear. In Fig. \ref{fig:wk}, the spin dispersion relations are displayed for ${J}_{2}/{J}_{1}= 0.10, \ 0.25$ and $0.60$, respectively. It clearly indicates that the BEC occurs at ${k}_{0}=(0, 0)$ when ${J}_{2}/{J}_{1}=0.10$ and $0.60$. For ${J}_{2}/{J}_{1}=0.25$, the system lies in the spin liquid phase. Although the lowest point of the dispersion relation is also at ${k}_{0}=(0, 0)$, the first excited state is separated from the ground state with a finite value. Hence there is no BEC and the system is disordered with a finite gap.
\begin{figure}[htbp]
\centering
\setlength{\abovecaptionskip}{2pt}
\setlength{\belowcaptionskip}{4pt}
\includegraphics[angle=0, width=1.0 \columnwidth]{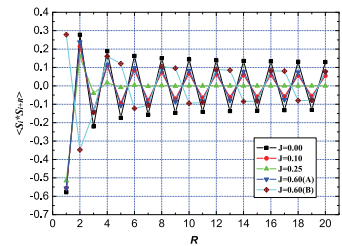}
\caption{(Color online) Spin-spin correlation function as a function of distance R in zigzag direction. For the N\'{e}el AFM phase ${J}_{2}/{J}_{1}=0$ and $0.1$; for $0.21<{J}_{2}/{J}_{1}<0.43$, the SSCF is short range,  indicating a gap zone with a short-range order; and for ${J}_{2}/{J}_{1}>0.43$, the correlations correspond to the collinear SAFM phase ($J=0.6(A)$ and $J=0.6(B)$ for two different zigzag directions).}
\label{fig:SSCF}
\end{figure}

The SSCF as the function of space distance of spins could disclose the spatial distribution of spin arrangement. For ${J}_{2}/{J}_{1} < 0.21$, the SSCF is N\'{e}el AFM coupling in each direction, and for ${J}_{2}/{J}_{1}>0.43$ the SAFM long-range correlations are also obtained in the zigzag direction, while the SSCF only shows short-range N\'{e}el AFM order for $0.21<{J}_{2}/{J}_{1}<0.43$ in the quantum disordered state with a gap in the bosonic dispersion. A plot of the SSCF for ${J}_{2}/{J}_{1}=0.0$, $0.1$, $0.25$ and $0.6$ in the zigzag direction is shown in Fig. \ref{fig:SSCF}. Although Tao $et\ al.$ pointed out that the quantitative features of the present SBMFT give the SSCF 1.5 times larger than that of the exact result\cite{PRB50.6840}, it is qualitatively correct. For ${J}_{2}/{J}_{1}=0.0$, our results divided by 1.5 are in exact agreement with the exact diagonalization and coupled-cluster method results given by Farnell\cite{PRB84.012403}. For ${J}_{2}/{J}_{1}=0.0$ and $0.1$, the SSCFs are isotropic and shows a classical N\'{e}el AFM order behavior. For ${J}_{2}/{J}_{1}=0.6$, the SSCFs are different along two different zigzag directions due to the SAFM structure, and the long-range correlations are obtained. When ${J}_{2}/{J}_{1}=0.25$, the SSCF rapidly decays to zero. Simultaneously, we notice that there is no symmetry breaking, so this state is a spin liquid state, which can be viewed as a superposition of short-range valence bond, also called a resonating valence bond (RVB) state. \\

\section{Finite-Temperature Properties of Spin liquid Phase}

At finite $T$, the N\'{e}el AFM and the SAFM phase become disordered phase due to strong gapless spin fluctuations, as constrained by the Mermin-Wagner theorem for the present two-dimensional system. While the gapped spin liquid phase could survive at finite $T$ until up to a critical temperature. Thus in the following we are especially interested in the finite-temperature properties of the spin liquid phase on the present spin frustrated honeycomb lattice. The nearest-neighbor spins are paired to form spin singlets, which are destroyed by the rising temperature gradually. Therefore, we treat ${J}_{1}$, corresponding to the strength of singlet, as a variable, and let ${J}_{2}$ equal to unity at finite temperatures.

\subsection{Order Parameters $\And$ Energy Gap}
\begin{figure}[htbp]
\centering
\setlength{\abovecaptionskip}{2pt}
\setlength{\belowcaptionskip}{4pt}
\includegraphics[angle=0, width=1.0 \columnwidth]{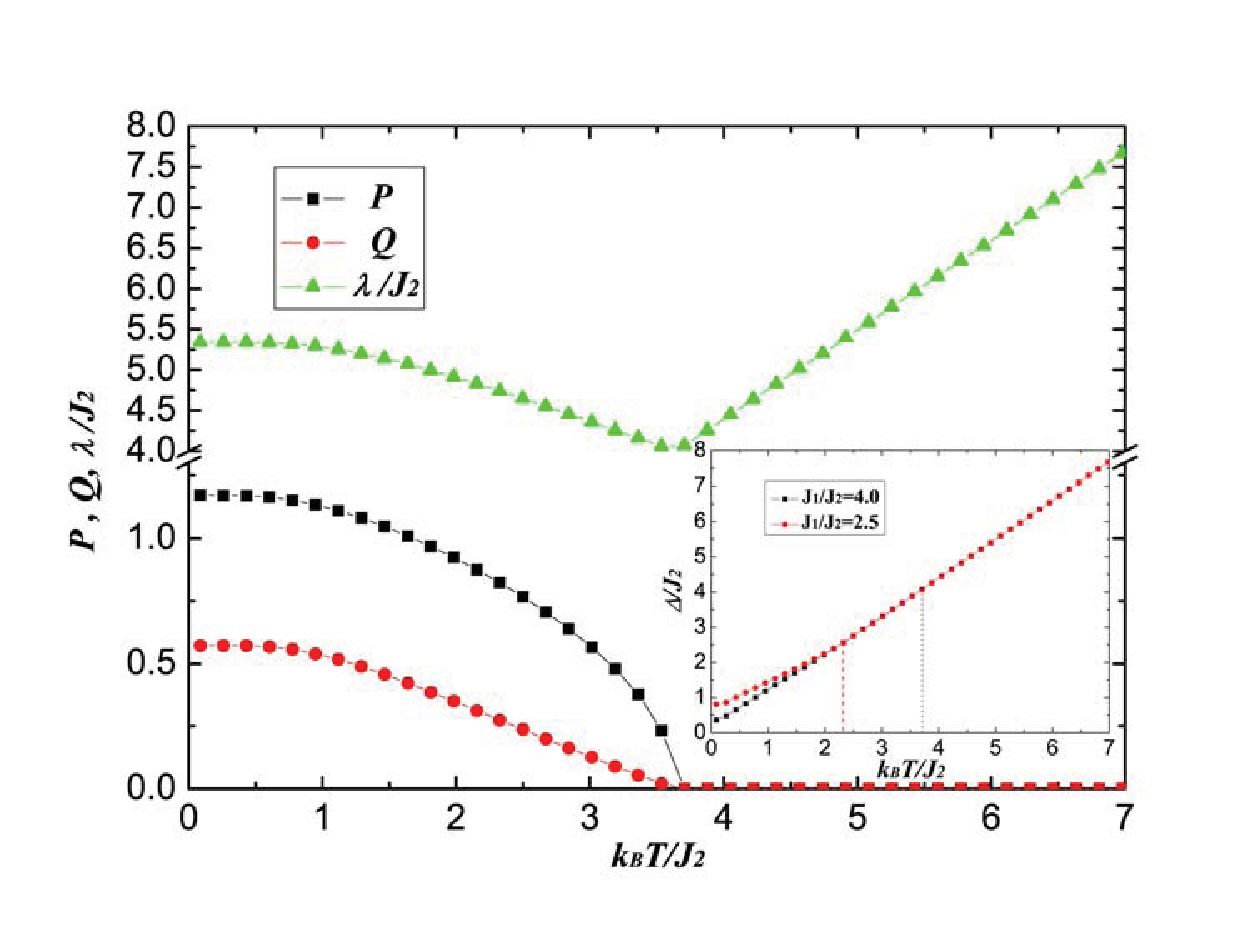}
\caption{(Color online) Temperature dependence of the RVB order parameter ${P}$, ${Q}$ and Lagrane multiplier $\lambda $ for $J_{1}/J_{2}=4.0$. Inset: energy gaps with different $J_{1}/J_{2}=4.0$ and $2.5$.}
\label{fig:P_Q_GAP}
\end{figure}

The variations of the order parameter $P, Q$ and Lagrange multiplier $\lambda $ with $T$ are obtained by solving the mean-field equations for $J_{1}/J_{2}=4.0$. The numerical results are shown in Fig. \ref{fig:P_Q_GAP}. With the increase of the temperature, the spin singlet correlations are suppressed by thermal fluctuations gradually. When the temperature is increased to about ${{k}_{B}}{{T}_{s}}/{J}_{2}=3.7$, both of the order parameters $P$ and $Q$ become zero simultaneously. This shows that strong thermal fluctuations in ${T}>{{T}_{s}}$ break the spin singlet pairs, and the system transits from the quantum disorder phase to the paramagnetic phase.

Temperature dependence of the energy gap is also presented in the inset of Fig. \ref{fig:P_Q_GAP}. The energy gap, which mainly arises from the contribution of the Lagrange multiplier, almost develops linearly with increasing temperature. Since in the finite and high temperatures, the Lagrange multiplier is an effective "chemical potential" of the Schwinger bosons; it is considerably larger than the energy scales of the ordered energies $J_{1}P$ and $J_{2}Q$, and almost linearly increases with the temperature. This results in the linear $T$ behavior of the energy gap when the system enters in the paramagnetic phase.

\begin{figure}[htbp]
\centering
\setlength{\abovecaptionskip}{2pt}
\setlength{\belowcaptionskip}{4pt}
\includegraphics[angle=0, width=1.0 \columnwidth]{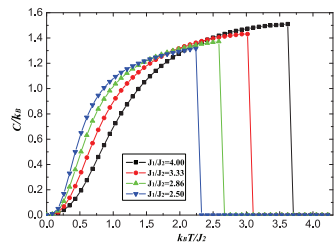}
\caption{((Color online) Temperature dependence of specific heat per site with different ${J}_{1}/{J}_{2}$ in the spin liquid phase.}
\label{fig:specific heat}
\end{figure}

\subsection{Specific Heat}

In finite temperatures, we calculate the specific heat of the spin liquid phase. The shape of $C-T$ is nearly like a $\lambda $-type transition, which is presented in Fig. \ref{fig:specific heat}.

When ${{k}_{B}}T/J_{2}=0$, the specific heat is zero since there is no any quasiparticle excitation. In the low-temperature region, an analytic form of the low-temperature specific heat per site is obtained
\begin{equation}
c_{v}\simeq \frac{{{S}_{c}}{{\Delta }^{2}}}{2\pi \eta T}{{e}^{-\Delta /{{k}_{B}}T}},\,
\end{equation}
where ${S}_{c}$ is the area of the unit cell of a sublattice, the energy gap is
\begin{equation}
\Delta =\sqrt{{{\left( 3{{J}_{2}}Q\text{+}\lambda  \right)}^{2}}-{{\left( \frac{3}{2}{{J}_{1}}P \right)}^{2}}},
\label{eq:delta}
\end{equation}
and the constant $\eta$ is
\begin{equation}
\eta =\frac{9J_{1}^{2}{{P}^{2}}-36{{J}_{2}}Q\left( 3{{J}_{2}}Q\text{+}\lambda  \right)}{\text{16}\sqrt{{{\left( 3{{J}_{2}}Q\text{+}\lambda  \right)}^{2}}-{{\left( \frac{3}{2}{{J}_{1}}P \right)}^{2}}}}.
\label{eq:eta}
\end{equation}
Here $P, Q$ and $\lambda$ are also constants when $T$ approaches to zero temperature. Such a low temperature behavior of spin liquid on a honeycomb lattice resembles that on a triangular lattice\cite{JPSJ60.614}.

When ${T}$ lifts up to ${{T}_{s}}$, the increasing temperature leads to more quasiparticle excitations, the contribution of the quasiparticle excitations to the specific heat becomes larger and larger. Meanwhile, the rising temperature destroys spin singlet. When $T$ is high enough to $T_{s}$, the spin singlets vanish and the phase transition from spin liquid to paramagnetic phase occurs, as shown in Fig. \ref{fig:specific heat}. Different ${J}_{2}/{J}_{1}$ corresponds to different strengths of spin singlet, leading to different critical temperatures. When the system transits from a RVB spin liquid state to a paramagnetic state, the specific heat sharply drops to zero, corresponding to a second-order phase transition similar to that seen in a square lattice\cite{JPSJ58.3733}.

\subsection{Magnetic Susceptibility}

\begin{figure}[htbp]
\centering
\setlength{\abovecaptionskip}{2pt}
\setlength{\belowcaptionskip}{4pt}
\includegraphics[angle=0, width=1.0 \columnwidth]{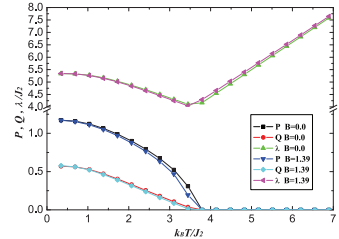}
\caption{Temperature dependence of the order parameters $P$, $Q$ and Lagrange multiplier $\lambda $ in the magnetic field $B=0$ and $3$, respectively. In the following figures, $B$ is measured in unit of $\mu B/{J}_{2}$.}
\label{fig:parameters}
\end{figure}

Under different magnetic fields, we get the temperature dependence of the RVB order parameter and Lagrange multiplier, which is presented in Fig. \ref{fig:parameters} for ${J}_{1}/{J}_{2}=4.0$. Throughout this paper, we keep external magnetic field $\mu B/{J}_{2} \le 1.39$, actually, which is not a small value. When $\mu B/{J}_{2}=1.39$, we obtain $B=60$ T by putting the value of $J_{2}=5$ meV. We find that the influence of the magnetic field on the order parameters is rather small, suggesting the spin liquid phase is robust and self-protected to applied magnetic field, as seen in Fig. \ref{fig:parameters}. This arises from that the spin singlet in the RVB spin liquid phase is rigid with respect to external magnetic field $B$ until a critical pair-breaking field $B_{c}$.

\begin{figure}[htbp]
\centering
\setlength{\abovecaptionskip}{2pt}
\setlength{\belowcaptionskip}{4pt}
\includegraphics[angle=0, width=1.0 \columnwidth]{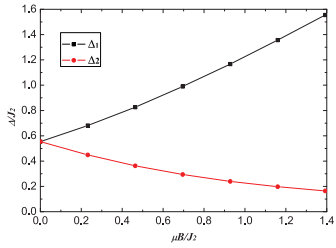}
\caption{(Color online) The separation of two energy gaps at ${{k}_{B}}T/{J}_{2}=0.345$ for ${J}_{1}/{J}_{2}=4.0$.}
\label{fig:gap-b}
\end{figure}

In finite magnetic fields, the energy dispersion of the Schwinger bosons splits into two branches, hence two energy gaps. The magnetic field dependence of the energy gaps, corresponding to Eq. (\ref{eq:omega_B}) at $k=(0,0)$, is shown in Fig. \ref{fig:gap-b} for ${J}_{1}/{J}_{2}=4.0$ and ${{k}_{B}}T/{J}_{2}=0.345$. With the increase of the magnetic field, two energy gaps go in different ways, one gap lifts and the other decreases. The separation between the two gaps linearly increases with applied magnetic fields, about energy scale of ${\mu }B$.

\begin{figure}[htbp]
\centering
\setlength{\abovecaptionskip}{2pt}
\setlength{\belowcaptionskip}{4pt}
\includegraphics[angle=0, width=1.0 \columnwidth]{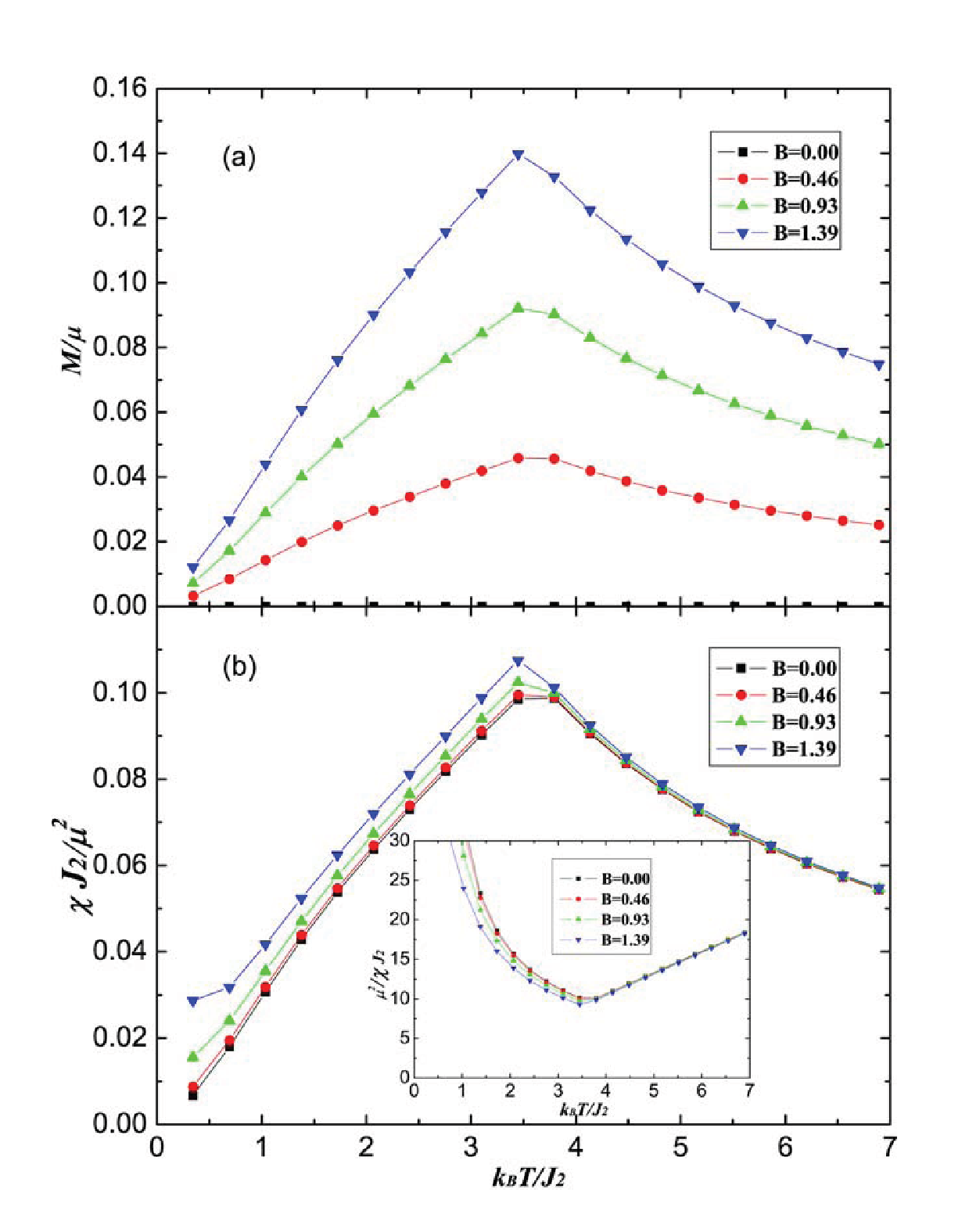}
\caption{(Color online) Temperature dependence of magnetization (a) and susceptibility (b) per site in different magnetic field B for ${J}_{1}/{J}_{2}=4.0$. Inset in (b): the reciprocal of the susceptibility (c) in different magnetic field B for ${J}_{1}/{J}_{2}=4.0$.}
\label{fig:susceptibility}
\end{figure}

We also investigate the temperature dependence of magnetization and susceptibility under different magnetic fields for ${J}_{1}/{J}_{2}=4.0$. The low-temperature behavior of the magnetic susceptibility per site is given as follows:
\begin{equation}
\chi \left( T \right)\simeq \frac{{{\mu }^{2}}{{S}_{c}}}{8\pi \eta }{{\operatorname{e}}^{{-\Delta }/{{{k}_{B}}T}\;}}\cosh \frac{\mu B}{2{{k}_{B}}T},
\end{equation}
where the energy gap $\Delta$ and constant $\eta$ are expressed in Eq. (\ref{eq:delta}) and (\ref{eq:eta}), respectively. When $B$ approaches to $0$, we get the magnetic susceptibility analytically
\begin{equation}
\chi \simeq \frac{{{\mu }^{2}}{{S}_{c}}}{\text{8}\pi \eta }{{\operatorname{e}}^{-\Delta /{{k}_{B}}T}},
\label{eq:low x}
\end{equation}
These low-T behaviors on a honeycomb lattice are very similar to Mila {\it et al.}'s on a square lattice\cite{PRB43.7891}.
There are three temperature ranges in Fig. \ref{fig:susceptibility}. Firstly, in extremely low temperature an exponential behavior of magnetic susceptibility as expressed in Eq. (\ref{eq:low x}). In the second temperature range, the magnetic susceptibility almost increases linearly with the lift of the temperature. In $0<T<{T}_{s}$, the spin singlet formed between two neighboring spins is gradually destroyed by the thermal fluctuations. Hence the spin is easier to be polarized so that the magnetization increases with the lift of temperature. Since the magnetic susceptibility characterizes the degree of difficulty for polarization and magnetization, it also becomes large for higher temperature when $T<{T}_{s}$. Finally, for $T>{T}_{s}$, the system transits from the spin liquid phase to the paramagnetic one. With the increase of the temperature, the thermal fluctuations of spins become more and more intense, depressing the spins polarization. Therefore, the magnetization and magnetic susceptibility gradually decrease in the high temperature region. Moreover, the inverse susceptibility is linearly dependent on temperature and consistent with the Curie law. Similar behavior is also obtained by Mila {\it et al.} for square lattice\cite{EJP21.499}, demonstrating that the SBMFT also works well in  high temperature $T>{T}_{s}$.

One may notice that such a paramagnetic phase with the Curie law is completely different from the conventional paramagnetic phase with the Curie-Weiss law. The quantum spin fluctuations in the former leads to vanishing static magnetic field on central spin, hence to the Curie law in magnetic susceptibility, suggesting a quantum paramagnetic phase. For the latter a conventional paramagnet with AFM coupling, the presence of the short range
coupling and effective magnetic field lead to a Curie-Weiss law.

\subsection{Knight Shift}

The low-temperature behavior of the Knight shift per site is given by
\begin{equation}
\kappa =\frac{K}{N}\simeq \frac{F(0)}{{{\gamma }_{N}}\hbar }\cdot \frac{{{S}_{c}}{{k}_{B}}T}{4\pi \eta B}{{\operatorname{e}}^{-\Delta /{{k}_{B}}T}}\sinh \frac{\mu B}{2{{k}_{B}}T},
\end{equation}
where the energy gap $\Delta$ and constant $\eta$ are expressed in Eq. (\ref{eq:delta}) and (\ref{eq:eta}), respectively.
When $B$ approaches to $0$, its analytic form reads
\begin{equation}
\kappa \simeq \frac{F(0)}{{{\gamma }_{N}}\hbar }\cdot \frac{\mu {{S}_{c}}}{8\pi \eta }{{\operatorname{e}}^{-\Delta /{{k}_{B}}T}}.
\end{equation}
Namely the Knight shift is exponentially small at $T\to 0$.

\begin{figure}[htbp]
\centering
\setlength{\abovecaptionskip}{2pt}
\setlength{\belowcaptionskip}{4pt}
\includegraphics[angle=0, width=1.0 \columnwidth]{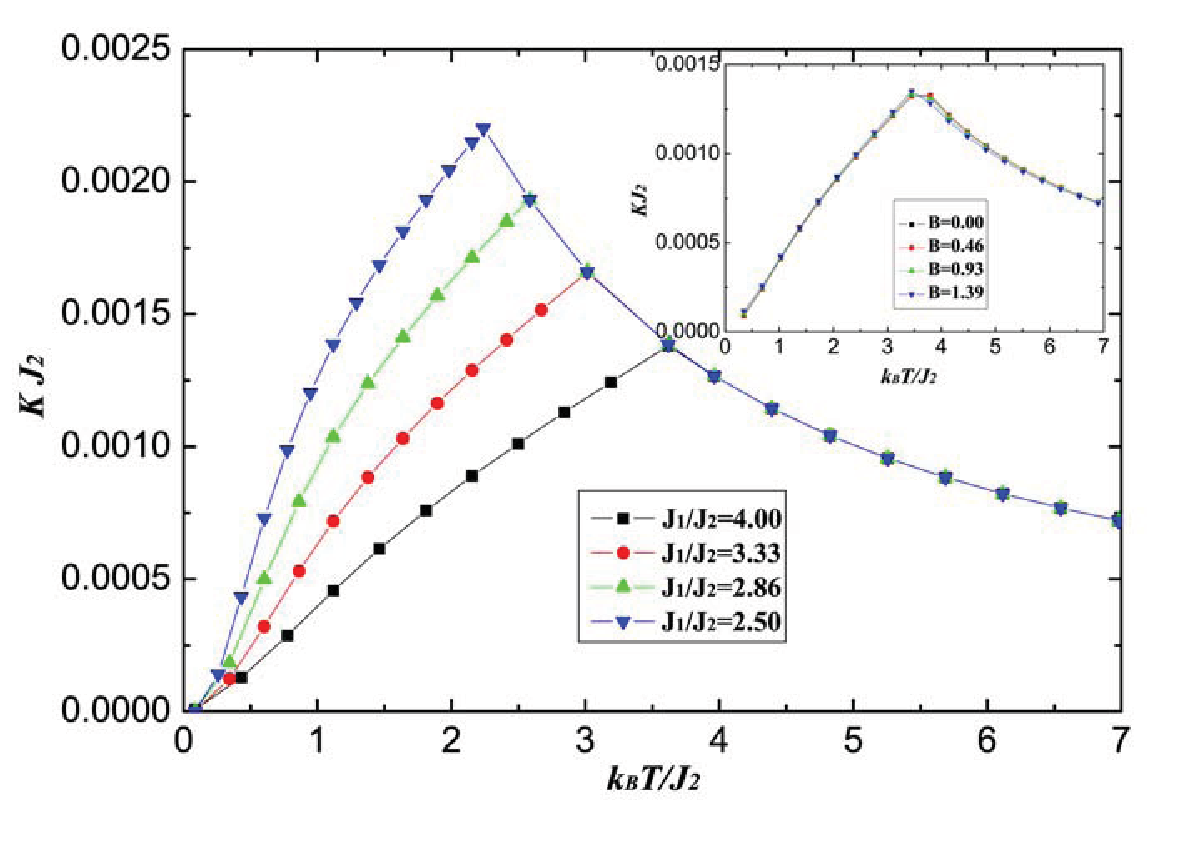}
\caption{(Color online) Temperature dependence of Knight shift per site for different ${J}_{1}/{J}_{2}$ at $\mu B/{J}_{2}=0.46$. The unit of $K$ is $ {F(0)}/{({{\gamma }_{N}}\hbar \mu )}\ $. Inset: temperature dependence of Knight shift per site in different magnetic field for ${J}_{1}/{J}_{2}=4.0$.}
\label{fig:Knight shift-B-J}
\end{figure}

The temperature dependence of the Knight shift for different ${J}_{1}/{J}_{2}$ at $\mu B/{J}_{2}=0.46$ is shown in Fig. \ref{fig:Knight shift-B-J}. With the increase of temperature, the Knight shift lifts with temperature with different slope for different ratio ${J}_{1}/{J}_{2}$ in the spin liquid phases.
It is interestingly found that the smaller the ratio ${J}_{1}/{J}_{2}$ is, the larger the Knight shift is in the spin liquid phase. This is because the strength of spin singlet is weak for small ${J}_{1}/{J}_{2}$, and the electron spin can be easily polarized by applied magnetic field, leading to a large Knight shift.
Once the spin liquid-paramagnetic phase transition occurs, all of the T-dependent Knight shifts fall into a single curve in the paramagnetic phase. These behaviors could be easily understood as follows:
for ${T}<{T}_{s}$, the rising temperature destroys the spin singlet and the AFM spin correlations become weak gradually. Accordingly, the electron spin is easier to be polarized so that the effective magnetic field around nuclear becomes larger. Thus the nuclear magnetic resonance (NMR) frequency and the Knight shift lift with the increasing temperature in ${T}<{T}_{s}$. When ${T}>{T}_{s}$, the system enters the paramagnetic phase, the intense thermal fluctuation considerably reduces the effective magnetic field around a nucleus. Hence, the Knight shift decreases with the increase of temperature in the paramagnetic phase.

We also present the magnetic field influence on the Knight shift of spin liquid phase. The temperature dependence of Knight-shift for different magnetic field at ${J}_{1}/{J}_{2}=4.0$ is shown in the inset of Fig. \ref{fig:Knight shift-B-J}. We find that the Knight-shift almost does not change when $\mu B$ varies from zero to $1.39{J}_{2}$. This is attributed to the fact that the effect of magnetic field on the order parameters $P$ and $Q$ is small, as seen in Fig. \ref{fig:parameters}, which leads to the spin singlet almost unchanged. Thus the Knight-shift in different magnetic fields changes very little.
From Fig. \ref{fig:Knight shift-B-J} and the inset one notices that the critical temperatures obtained from the Knight shift are also in agreement with those obtained from the specific heat in Fig. \ref{fig:specific heat}.

\section{REMARKS}

From the preceding study we have found that through comparing the total energy, there are three stable phases in the phase diagram of frustrated ${J}_{1}-{J}_{2}$ Heisenberg model on a honeycomb lattice: the N\'{e}el AFM phase in ${J}_{2}/{J}_{1}<0.21$, the RVB spin liquid phase in $0.21<{J}_{2}/{J}_{1}<0.43$, and the collinear SAFM phase in ${J}_{2}/{J}_{1}>0.43$ . Although several authors suggested a plaquette valence bond crystal state in the intermediate frustration regime, instead of the RVB spin liquid one, through the exact diagonalization method \cite{JPCM23.226006, PRB84.024406, PRB84.012403} and coupled-cluster method\cite{JPCM24.236002}, it was not known whether the finite-size result is stable in the thermodynamic limit ( $i.e.$ $N \approx infty$ ). We also notice that the critical value of the spin liquid-N\'{e}el AFM boundary in the Hubbard model\cite{PRL107.087204} is smaller than that of ours, this is attributed to the duplicate effects of the spin fluctuations and the charge fluctuations, which prevent the formation of magnetic long-range order.

We find that beyond the phase transition temperature ${T}_{s}$, the magnetic susceptibility recovers the usual Curie behavior. These results demonstrated that the Schwinger-boson mean-field theory works well for calculating thermodynamic properties of the frustrated ${J}_{1}-{J}_{2}$ Heisenberg model on a honeycomb lattice, even if in the high temperature regime without magnetic order. This arises from the finite value of the gap parameters $\lambda$, even if the order parameters $P$ and $Q$ vanish in the high temperature. More importantly, we present the low-temperature exponential behavior of the specific heat, magnetic susceptibility and Knight shift, providing a clue for searching the candidate compounds with exotic spin liquid phase in experiments. We also illustrate that it is not easy for an external magnetic field destroy the spin liquid phase, showing a self-protecting and robust characteristics of this quantum phase.

\acknowledgements
One of the authors (L. J. Zou) thanks useful comments and discussions with S.-Q. Shen.
This work was supported by the National Science Foundation of China
under Grant No.11074257, 11074177, and 11274310, SRF for ROCS SEM (20111139-10-2),
Knowledge Innovation Program of Chinese Academy of Sciences, and Director Grants of CASHIPS.
Numerical calculations were performed in Center for Computational
Science of CASHIPS.


\end{document}